\documentclass[apj]{emulateapj}
\pdfoutput=1
\usepackage[colorlinks,linkcolor={blue},citecolor={blue},urlcolor={red}]{hyperref}
\usepackage{epsfig} 
\usepackage{graphicx}
\usepackage{natbib} 
\usepackage{amsmath} 
\usepackage{amsfonts}
\usepackage{amssymb}

\def\msun{\,{\rm h^{-1} M}_\odot}

\def\mpch{\,{\rm h^{-1}  Mpc}}

\def\bx{{\boldsymbol x}}

\def\e{\,{\boldsymbol{e_1}}}
\def\ee{\,{\boldsymbol{e_2}}}
\def\eee{\,{\boldsymbol{e_3}}}

\begin{document}

\title{Alignment between satellite and central galaxies in the SDSS DR7: dependence on large-scale environment}

\author{Peng Wang\altaffilmark{1,2,3},
Yu Luo\altaffilmark{1},
Xi Kang\altaffilmark{1},
Noam I. Libeskind\altaffilmark{2},
Lei Wang\altaffilmark{1},
Youcai Zhang\altaffilmark{4},
Elmo Tempel\altaffilmark{2,5},
Quan Guo\altaffilmark{6}}

\altaffiltext{1}{Purple Mountain Observatory (PMO), 2 West Beijing Road, Nanjing 210008, China}
\altaffiltext{2}{Leibniz-Institut f\"ur Astrophysik Potsdam (AIP),  An der Sternwarte 16, 14482 Potsdam, Germany}   
\altaffiltext{3}{Graduate School, University of the Chinese Academy of Science, 19A, Yuquan Road, Beijing 100049, China}
\altaffiltext{4}{Key Laboratory for Research in Galaxies and Cosmology, Shanghai Astronomical Observatory (SHAO), Nandan Road 80, Shanghai 200030, China}
\altaffiltext{5}{Tartu Observatory, University of Tartu, Observatooriumi 1, 61602 T\~oravere, Estonia}
\altaffiltext{6}{Shanghai Astronomical Observatory (SHAO), Nandan Road 80, Shanghai 200030, China}

\email{wangpeng@pmo.ac.cn(PW)}
\email{luoyu@pmo.ac.cn(YL)}
\email{kangxi@pmo.ac.cn(XK)}


\begin{abstract}

The alignment between satellites and central galaxies has been studied in detail both in observational and theoretical works. The widely accepted fact is that the satellites preferentially reside along the major axis of their central galaxy. However, the origin and large-scale environment dependence of this alignment are still unknown. In an attempt to figure out those, we use data constructed from SDSS DR7 to investigate the large-scale environmental dependence of this alignment with emphasis on examining the alignments' dependence on the colour of the central galaxy. We find a very strong large-scale environmental dependence of the satellite-central alignment in  groups with blue centrals. Satellites of blue centrals in knots are preferentially located perpendicular to the major axis of the centrals, and the alignment angle decreases with environment namely when going from knots to voids. The alignment angle strongly depend on the ${}^{0.1}(g-r)$ colour of centrals.  We suggest that the satellite-central alignment is the result of a competition between  satellite accretion within large scale-structure and galaxy evolution inside host haloes. For groups containing red central galaxies, the satellite-central alignment is mainly determined by the evolution effect, while for blue central dominated groups, the effect of large-scale structure plays a more important role, especially in knots. Our results provide an explanation for how the satellite-central alignment forms within different large-scale environments. The perpendicular case in groups and knots with blue centrals may also provide insight into understanding similar polar arrangements such the formation of the Milky Way and Centaurus A's satellite system.

\end{abstract}

\section{Introduction}
\label{sec:intro}
In the current standard cold dark matter (CDM) cosmological model, haloes (and the galaxies embedded inside of them) are assembled hierarchically by smaller structures predominantly though mergers \citep{1978MNRAS.183..341W}. Density perturbations collapse and form virilalzed dark matter haloes which are then accreted by successively larger, gravitationally bound structures. Once inside a larger halo, accreted dark matter haloes are termed ``sub-haloes'' and their galaxies called ``satellites''. The exact geometry and nature of halo growth via accretion and mergers is therefore determined by the large scale structure (hereafter LSS). Understanding how large dark matter haloes grow, implies understanding how the LSS feeds sub-haloes and dark matter towards them.

Since sub-haloes (satellite galaxies) are mainly distributed throughout the whole host halo (galaxies), they are ideally selected as  tracers to study the mass distribution in haloes. Recently theoretical works \citep{2011MNRAS.411.1525L, 2014MNRAS.443.1274L, 2015ApJ...807...37S, 2015ApJ...813....6K, 2018MNRAS.473.1562W, 2018MNRAS.476.1796S} have shown that the sub-haloes (satellites) are preferentially accreted along the axis of the LSS (see also \cite{2017MNRAS.472.4099K}  for non-CDM accretion alignment). More and more studies \citep[][]{2007ApJ...655L...5A, 2014MNRAS.440L..46A, 2007MNRAS.381...41H, 2007MNRAS.375..489H, 2010MNRAS.405..274H, 2009ApJ...706..747Z, 2015ApJ...798...17Z, 2017MNRAS.468L.123W, 2018MNRAS.473.1562W, 2015MNRAS.446.1458M, 2012MNRAS.421L.137L, 2013MNRAS.428.2489L, 2013ApJ...766L..15L, 2016MNRAS.457..695P, 2015MNRAS.450.2727T, 2015ApJ...800..112G, 2013ApJ...775L..42T, 2014MNRAS.438.3465T, 2017ApJ...848...22X, 2017A&A...599A..31H} suggested that haloes (galaxies) properties (such as shape and spin) are affected by the LSS. Consequently, the sub-haloes (satellites) also can be used as a bridge between the properties of haloes (galaxies) and LSS.

\cite{1968PASP...80..252S} was the first to examine the question of alignments in extra galactic observations. Using a relatively small sample of galaxies in knots, he reported an alignment between the position vector of satellites and the orientation of their host galaxy's major axis. One year later, \cite{1969ArA.....5..305H} studied the distribution of satellite galaxies around isolated disc galaxies (within  60 $\rm kpc$) and found the opposite: that the satellites tend to be preferentially perpendicular to the major axis of the central galaxy. This was coined the $Holmberg$ effect and may be related to alignments seen in the local universe
\citep{2015ApJ...802L..25T, 2012MNRAS.423.1109P, 2013MNRAS.435.1928P, 2016MNRAS.460.3772S, 2018arXiv180200081M},
although these tend to be seen on larger scales than those originally probed by Holmberg \citep[see also][]{1997ApJ...478L..53Z}. However, the picture appears reversed wthin a number of studies which look at large galaxy surveys, such as the Sloan Digital Sky Survey (SDSS). \cite{2005ApJ...628L.101B} and \cite{2006MNRAS.369.1293Y} (hereafter Y06) obtained a unified and widely accepted conclusion by using a large sample of galaxy groups selected from the SDSS DR4, which suggested that the satellites tend to align with the major axis of the central galaxy, and the alignment depends on galaxy properties, such as the colour of galaxies and the mass of the groups. The alignment signal is strongest between red centrals and red satellites in massive groups, and almost absent between blue centrals and blue satellites in low-mass groups.

To understand the observed alignment signals, theoretical works have progressed in tandem. \cite{2006ApJ...650..550A} and \cite{2007MNRAS.378.1531K} reproduced the alignment seen in large surveys using N-body simulation where galaxies were modelled semi-analytically.  Similarly, \cite{2005MNRAS.363..146L},  \cite{2007MNRAS.374...16L}, \cite{2005A&A...437..383K, 2007MNRAS.378.1531K} and \cite{2005ApJ...629..219Z} were able to somewhat explain ``great disks'' such as those seen in the local Universe. However, these works typically show a stronger alignment than observed, due to simply to the  assumption that the shape of the central galaxy follows the shape of its host halo without considering the mis-alignment between the central galaxy and host dark matter halo \citep[see][]{2007MNRAS.374...16L, 2009RAA.....9...41F, 2009ApJ...694..214O, 2011MNRAS.415.2607D, 2015MNRAS.453..721V, 2015ApJ...800...34G, 2014ApJ...791L..33D}. \cite{2014ApJ...786....8W} claimed there is a weaker alignment which agrees with observational data if the central galaxy has the same shape as the inner region of host halo.

Although observations of the satellite-central alignment (hereafter SCA) have been clearly confirmed both in simulations and observations,  their origin is still controversial. An immediate question is why the SCA depends on galaxy properties, such as central or satellite galaxy colour. Is this attributed to primordial infall or an evolutionary effect inside the halo? In another words, is the alignment nature or nurture?  Moreover, the infalling satellite galaxies lie in different LSS, so what is the impact of LSS on the SCA? With the exception of \cite{2005MNRAS.363..146L}, who found an alignment between the satellite planes seen in the Local Volume and the local shear field such considerations are mostly ignored in most studies.  This leads to other related questions: is the SCA affected by the LSS?  if so, how?  The main goal of this paper is to look for a possible explanation of these two questions using the SDSS DR7.

The paper is organized as follows: we first present the group catalogue, definitions of large-scale structure and the three different alignment angles in Section 2. Section 3 contains the main results of the paper by comparing the satellites-central alignment and its large-scale environment dependence. We will discuss and summarize our results in Section 4.

\section{Data and Methodology}
The observational data used here are taken from the SDSS DR7. In this section, we give a brief description of data, and how the cosmic web is characterised. We refer the readers to \cite{2006MNRAS.369.1293Y}, \cite{2012MNRAS.420.1809W}  and  \cite{2013ApJ...779..160Z} for more details.

\subsection{Group catalogue}
The galaxy catalogue is constructed from the New York University Value-Addad Galaxy Catalog \citep[NYU-VAGC,][]{2005AJ....129.2562B}, which is based on the SDSS Data Release 7 \citep[DR7][]{2009ApJS..182..543A}. The total number of galaxies in this catalogue is 639,555 with redshifts in the range $[0.01,0.20]$ and with magnitudes brighter than $r=17.72$. Galaxy groups are selected using an adaptive halo-based group finder which was developed by \cite{2005MNRAS.356.1293Y, 2007ApJ...671..153Y}. In each group, we refer  to the most massive galaxy (MMG) as the central, whereas all other members are considered satellites. We pre-emptively note that there is no significant change to our results if we choose the brightest galaxy as the central. In order to make robust results and reduce the uncertainty of the measurement of projected major axis of central galaxies, we follow \cite{2006MNRAS.369.1293Y} to use the central galaxies with ellipticity (minor to major axis ratio) $e\geq0.2$.
Galaxies are separated into sub-samples according to their ${}^{0.1}{(g-r)}$ colours, which corresponds to the $g-r$ colour k-corrected to redshift $z=0.1$.  We use the value of ${}^{0.1}{(g-r)}=0.83$ \citep[see][]{2006MNRAS.366....2W,2006MNRAS.369.1293Y} as the criterion for identifying galaxies are red or blue. Galaxies with ${}^{0.1}{(g-r)}>0.83$ are 'red', otherwise they are 'blue'. 
We divided our sample into 8 sub-samples: ``red centrals'', ``blue centrals'', ``red satellites'', ``blue satellites'', ``red-red'', ``red-blue'', ``blue-red'' and ``blue-blue''. The number of satellite-central pairs are list in the Table~\ref{table:table1}.
For ''red/blue centrals (satellites)'', we only consider the colour of centrals (satellites). For ``red-red (blue)'', we first select all red centrals and then pick out red (blue) satellites belong to those red centrals.

\subsection{Cosmic web}
The cosmic web catalogue of SDSS DR7 used in this work is calculated by \cite{2012MNRAS.420.1809W} using the tidal tensor field
\begin{equation}
T_{ij}(\bx) = \frac {\partial^2 \phi} {\partial x_i \partial x_j},
\label{TidalTensor}
\end{equation}
in which $i, j= 1,2,3$ are indices representing spatial dimensions.  $\phi$ is the peculiar gravitational potential calculated from the distribution of groups with mass greater than a threshold value $M_{th}=10^{12} \msun$ with a smoothing scale of $2.1 \mpch$. The details can be found in \cite{2009MNRAS.394..398W, 2012MNRAS.420.1809W} and \cite{2013ApJ...779..160Z}.

The tidal tensor is subject to a principle component analysis and its eigenvalues with corresponding eigenvectors are computed. The large scale environment (hereafter LSE)around a group/halo is defined using the eigenvalues (labeled in decreasing order $\lambda_{1}>\lambda_{2}>\lambda_{3}$), and it is classified as $knot$, $filament$, $wall$, or $void$ depending on the number of positive eigenvalues akin to the seminal work of Zel'dovich theory \citep{1970A&A.....5...84Z}. The trace of the tidal tensor, $\lambda_{1}+\lambda_{2}+\lambda_{3}$ scales as the density, on average the order of the mean density is $\rm Knot>Filament>Wall$ \citep[see Fig.4 in][for more detail]{2018MNRAS.473.1195L}
The eigenvector $\eee$ (corresponding to $\lambda_{3}$)  defines the direction of slowest collapse of mass on large scale. For example, in a filament environment, $\eee$ points along the spine of the filament, while in a wall environment, $\eee$ lies in the wall plane ($\ee$-$\eee$). Here $\e$ is the normal of the wall. Note that eigenvalues and corresponding eigenvectors are calculated at the position of each group in 3-D space at first, and then we project them to sky plane \citep[see Section 2.3 in][]{2013ApJ...779..160Z}, so in the following text, we use $\eee$ refers to the projected $\eee$.

\subsection{Alignment angles}
We mainly focus on the SCA, defined by the distribution of the alignment angle, $\theta_{\rm C-S}$. This can be quantified with the following method. Observationally, the angle is defined in the plane of the sky as the angle between the projected major axis of the central galaxy (which is determined by the 25-mag $\rm arcsec^{-2}$ isophote in the r band), and the direction of a satellite's position. Additionally, the alignment angle $\theta_{\rm C-\eee}$ is that between the major axis of central galaxy and the projected $\eee$ of tidal field. Similarly the angle $\theta_{\rm S-\eee}$ is the angle between the satellite position vector (relative to its central) and the projected $\eee$.

The alignment angles are restricted into the range $[0^{\circ},90^{\circ}]$, where $0^{\circ}$ implies that perfect parallel alignment and $90^{\circ}$ implies a perpendicular orientation. In other words if $\theta_{\rm C-S}=0^{\circ} (90^{\circ})$,  the satellites are aligned with (perpendicular to) the projected major axis of centrals. In the case of a random (uniform) distribution in the plane of the sky, the expected mean value of the distribution of angles is $\langle \theta \rangle = 45^{\circ}$. If $\langle \theta_{\rm C-S}\rangle <45^{\circ}$, we refer to the distributions as preferentially aligned while if $\langle\theta_{\rm C-S}\rangle>45^{\circ}$, it means satellites are preferentially residing in the regions perpendicular to the major axis of central.


\section{Results}
\label{sec:results}
In this section, we first repeat the previous work of Y06 regarding the colour and mass dependence of the SCA to see if we recover their results with the DR7. Then we check if there is any LSE dependence of the SCA. Finally, we try to understand the origin of the colour and LSE dependence of the SCA by investigating the radial distribution of the angle among satellites, central galaxies and $\eee$. The satellite radial distribution in different LSE may indicate not only the evolution of satellites in a group but also the interaction between satellites and central galaxies.

\begin{table}
\caption{The mean angle, $\theta_{\rm C-S}$, its error,  significance expected from a uniform distribution (in parentheses) and the sample size of the all sample and 8 sub-samples used in this work. Here, sub-samples are divided according to their colour. In row(2-3), we only consider the colour of central galaxies, whereas in row(4-5), the colour the satellites are only considered. In row(6-9), we selected central-satellites pairs according to the colours both central and satellite galaxies. Note that, the name of sub-samples in row(6-9), such as ``red-red'', the first ``red'' represents the colour of central, and second ``red'' for satellite.}
\begin{center}
\begin{tabular}{cccc}
\hline
 Sample Name        &  Y06                  &   This work  & No.  \\
\hline
all sample          &  $42.2\pm0.2 \ (14.0\sigma)$         & $42.2\pm0.06 \ (46.3\sigma)$ &118939 \\
\hline
red centrals         &  $41.5\pm0.2 \ (17.5\sigma)$         & $41.7\pm0.10 \ (33.0\sigma)$  & 92243 \\
blue centrals       &  $44.5\pm0.5 \ (1.0\sigma)$         & $44.7\pm0.15 \ (2.0\sigma)$ & 26696  \\
\hline
red satellites       &  $41.5\pm0.3 \ (11.7\sigma)$         & $41.5\pm0.11 \ (31.82\sigma)$   & 56797 \\
blue satellites     &  $43.3\pm0.3 \ (5.7\sigma)$         & $43.2\pm0.09 \ (20.00\sigma)$  & 62142 \\
\hline
red - red             &  $40.8\pm0.3 \ (14.0\sigma)$         & $40.9\pm0.12 \ (34.17\sigma)$  & 49563 \\
red - blue           &  $42.9\pm0.3 \ (7.0\sigma)$         & $42.6\pm0.13 \ (18.46\sigma)$   & 42680 \\
blue - red           &  $44.8\pm0.7 \ (0.3\sigma)$         & $45.5\pm0.31 \ (1.61\sigma)$   & 7234 \\
blue - blue         &  $44.2\pm0.6 \ (1.2\sigma)$         & $44.4\pm0.20 \ (3.00\sigma)$  &  19462 \\
\hline
\end{tabular}
\end{center}
\label{table:table1}
\end{table}

\begin{figure}
\centerline{\includegraphics[width=0.5\textwidth]{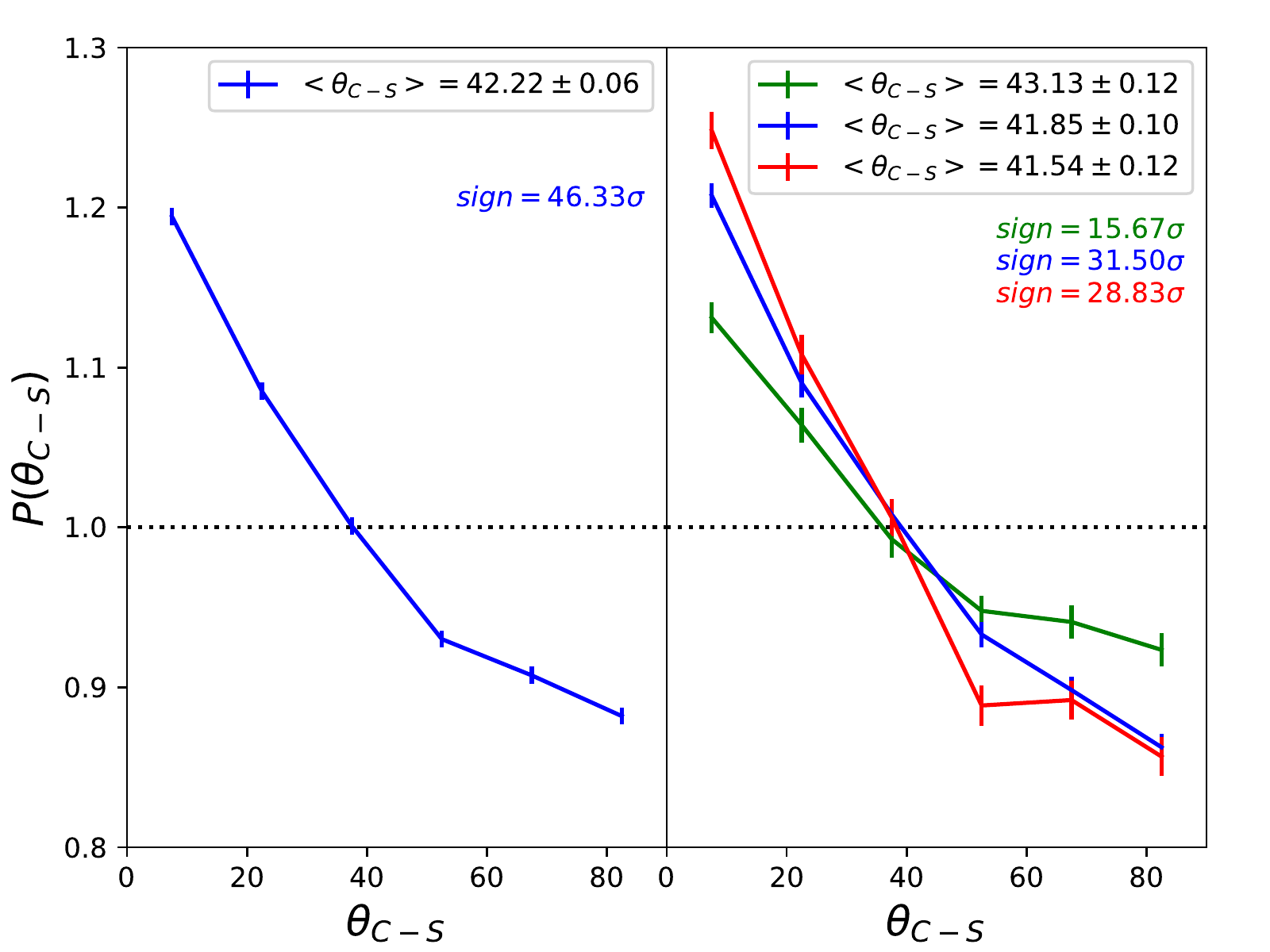}}
\caption{The normalized probability distribution of the angle $\theta_{\rm C-S}$ between the orientation of the projected major axis of the central galaxy and the position vector of each satellite point to the central galaxy in SDSS DR7. The normalized and error bars are computed from 100 random samples in which we have randomized the orientation of all central galaxies (see text for details). The average value of $\theta_{\rm C-S}$, its error and significance expected from a uniform distribution are indicated in the upper right-hand corner in each panel, respectively.  Note that an isotropic satellite distribution corresponds to $\rm PDF=1$ ($\langle\theta\rangle=45^{\circ}$). In the right panel, different colours represent different mass range with green showing halos with mass in the interval $[10^{12},10^{13}]$, blue in the interval $[10^{13},10^{14}]$ and  red in the interval $[10^{14},+\infty]$, in units of $\msun$.}
\label{fig:fig0}
\end{figure}

\begin{figure*}[t]
\centerline{\includegraphics[width=0.85\textwidth]{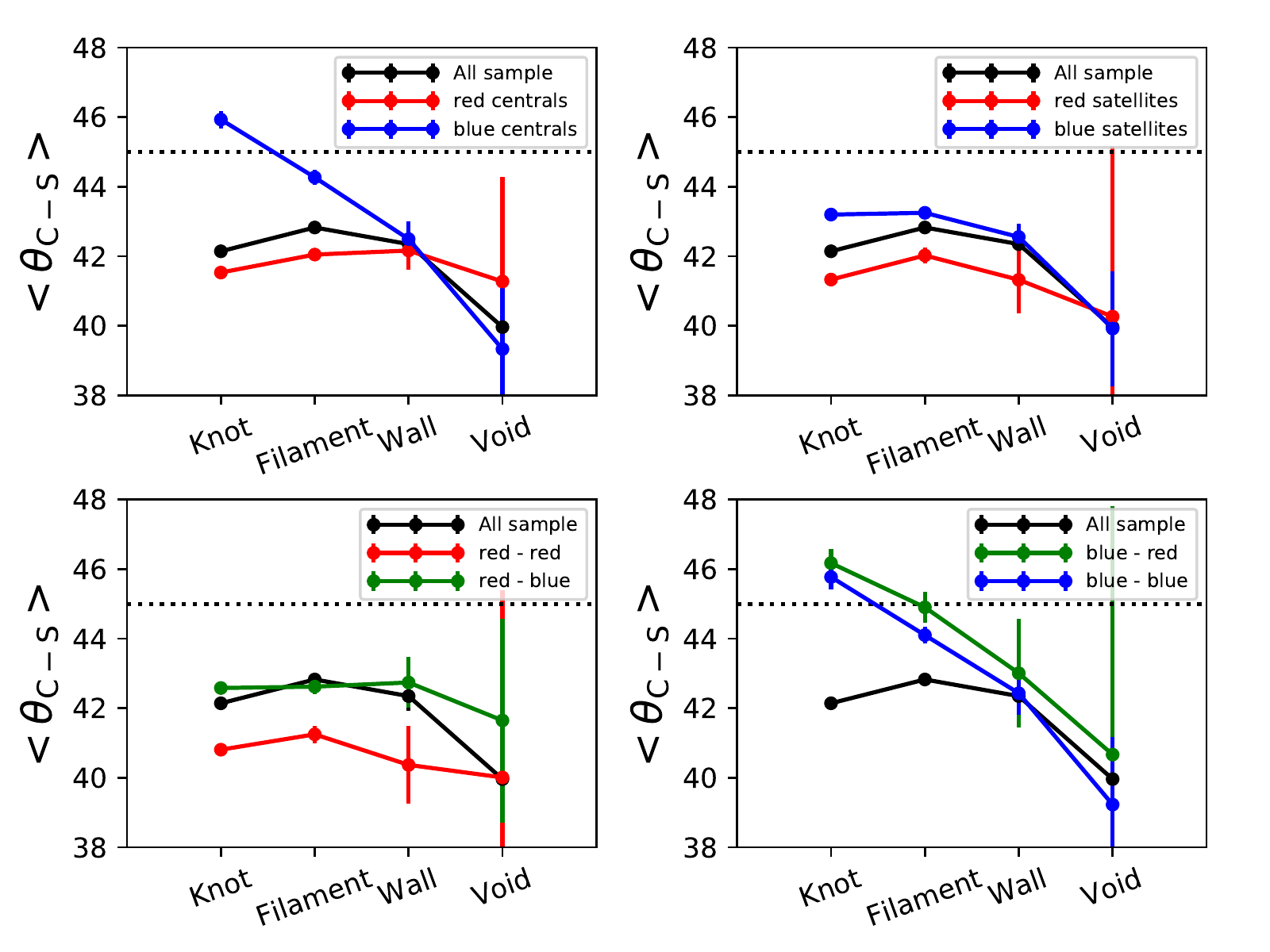}}
\caption{The mean angle of the SCA in different environments by considering colour dependence of satellites and centrals.  In each panel, we show the mean $\theta_{\rm C-S}$ and their error bars. The significance of each sample is shown in Table~\ref{table:table2}. The thin, dotted horizontal line indicates $\langle\theta_{\rm C-S}\rangle$=$45^{\circ}$ , which corresponds to an isotropic distribution. Back lines represents the all pairs in this work. In upper two panels, we only consider the colour of  centrals/satellites (same as row (2-3)/(4-5) in Table~\ref{table:table1}).  In bottom two panels, we selected sub-samples according to the colours of both the centrals and the satellites (same as row (6-9) in Table~\ref{table:table1}), as indicated.}
\label{fig:fig1}
\end{figure*}

\begin{table*}[t]
\caption{The mean angle $\rm \theta_{C_S}$ and its significance, which expected from a uniform distribution, for each sample in Fig.~\ref{fig:fig1}. }
\begin{center}
\begin{tabular}{ccccc}
\hline
                               &  Knot         &  Filament     &    Wall      &   Void  \\
\hline
all sample       & $42.1 \ (29.0\sigma)$ & $42.8 \ (18.0\sigma)$ & $42.3 \ (6.4\sigma)$ & $40.0 \ (3.1\sigma)$ \\
\hline
red centrals     & $41.5 \ (32.3\sigma)$ & $42.0 \ (18.9\sigma)$ & $42.2 \ (5.2\sigma)$ & $41.3 \ (1.2\sigma)$ \\
blue centrals    & $45.9 \ (3.7\sigma)$ & $44.3 \ (3.5\sigma)$ & $42.5 \ (4.9\sigma)$ & $39.3 \ (3.1\sigma)$ \\
\hline
red satellites   & $41.3 \ (28.3\sigma)$ & $42.0 \ (13.2\sigma)$ & $41.3 \ (3.8\sigma)$ & $40.3 \ (1.0\sigma)$ \\
blue satellites  & $43.2 \ (13.6\sigma)$ & $43.2 \ (10.2\sigma)$ & $42.6 \ (6.4\sigma)$ & $39.9 \ (3.1\sigma)$ \\
\hline
red - red        & $40.8 \ (32.2\sigma)$  & $41.2 \ (15.1\sigma)$ & $40.4 \ (4.1\sigma)$ & $40.0 \ (0.9\sigma)$ \\
red - blue       & $42.6 \ (16.7\sigma)$ & $42.6 \ (11.5\sigma)$ & $42.7 \ (3.1\sigma)$ & $41.6 \ (1.1\sigma)$ \\
blue - red       & $46.2 \ (2.8\sigma)$  & $44.9 \ (0.2\sigma)$ & $43.0 \ (1.3\sigma)$ & $40.7 \ (0.6\sigma)$ \\
blue - blue      & $45.8 \ (2.2\sigma)$  & $44.1 \ (3.6\sigma)$ & $42.4 \ (4.2\sigma)$ & $39.2 \ (3.0\sigma)$ \\
\hline
\end{tabular}
\end{center}
\label{table:table2}
\end{table*}

\subsection{The colour and mass dependence of the SCA}
We start our analysis while repeating the colour and mass dependence of the SCA and compare with Y06. We divided our central-satellite pairs into 8 different sub-samples according to the colours of central galaxies and satellites.  In Table~\ref{table:table1}, we list the results of mean $\theta_{\rm C-S}$ in Y06 and this work in each sub-sample.  By following the error calculation method in Y06, the error bars shown in Table~\ref{table:table1} (also in following figures) are computed from 100 random samples in which we have randomized the orientation of the projected major axis of all central galaxies (see Y06 for more details). Table~\ref{table:table1} clearly demonstrates that our results are fully consistent with Y06 albeit with much higher confidence levels ($46.3\sigma$) due to the larger samples selected from DR7 compared to Y06's SDSS DR4 ($14\sigma$). Just as Y06 did, we also show that the alignment signal is stronger for red satellites of red central galaxies. Again, similar to Y06 the alignment signal is nearly absent for blue satellites in the blue central groups. In Fig.~\ref{fig:fig0}, we show the probability distribution of the alignment angle, $\theta_{\rm C-S}$, between satellites and centrals. The mean values of $\theta_{\rm C-S}$ with errors and their significances are also shown at the top of each panel, respectively. The left panel shows the results for all samples, while the right panel shows the mass dependence considering group halo mass within three mass bins: $[10^{12},10^{13}]$, $[10^{13},10^{14}]$ and $[10^{14},+\infty]$ (green, blue and red, respectively), in units of $\msun$. Clearly, we see that there is a weak but significant mass dependence of the alignment: more massive groups have their satellites more aligned with the central's major axis. This finding is consistent with previous works \citep[][and reference therein]{2006MNRAS.369.1293Y, 2007MNRAS.378.1531K, 2014ApJ...786....8W}. Since the mass dependence is fairly weak, and the colour dependence still exists even for the same mass range of host halo \citep{2006MNRAS.369.1293Y},  we only focus on the colour dependence of the SCA in the following section.

\begin{figure*}
\centerline{\includegraphics[width=0.85\textwidth]{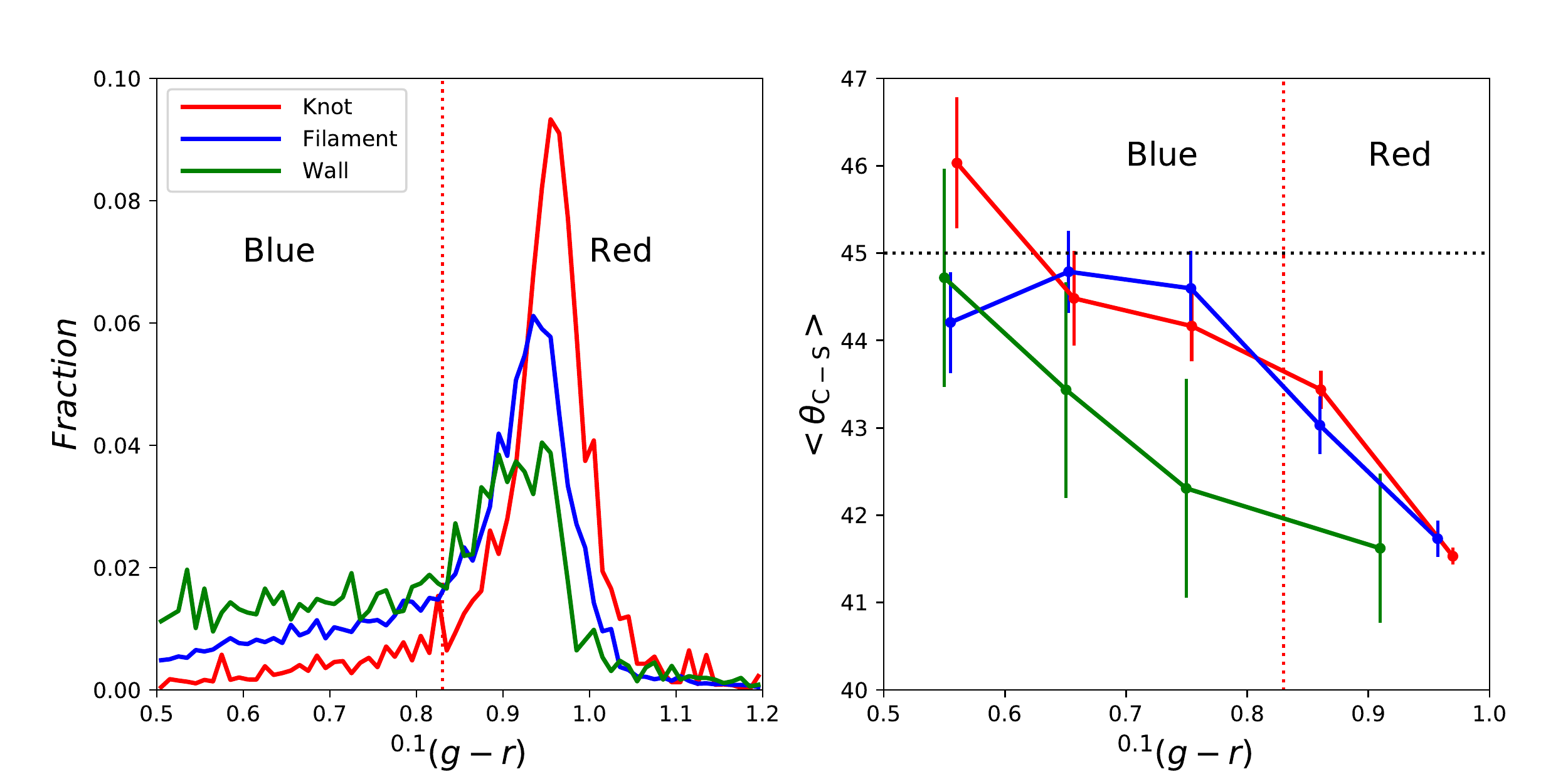}}
\caption{Left panel: the number fraction distribution of the colour of central galaxies in knots, filaments and walls (red, blue, and green curves respectively). Right panel: the $\theta_{\rm C-S}$ as function of colour of central galaxies. Here Poisson error are shown for each sample. The red dashed line in each panel indicates the criterion used to divided galaxies into ``red'' and ``blue''. The black dashed line in the right panel represents that the $\theta_{\rm C-S}$ is uniform aligned with ${}^{0.1}{(g-r)}$.}
\label{fig:fig2}
\end{figure*}

\subsection{The LSE effect on the SCA}
Conventionally colour is used as a proxy for the age of a stellar population, namely a redder galaxy is older and is more likely to lie in a dense region subjecting it to more merger events \citep{1980ApJ...236..351D}. Thus we expect a clear colour dependence of SCA to imply the formation history of both the central galaxies and satellites. Besides, their environment may play the role in SCA. In order to quantify the impact of the LSE on the SCA, we show the mean $\theta_{\rm C-S}$ as a function of the LSE (as defined by the tidal tensor) as well as considering the colour of central-satellite pairs, divided as in Table~\ref{table:table1}.  In Fig.~\ref{fig:fig1} we show the $\langle\theta_{C-S}\rangle$ as a function of LSE for each sub-sample, and their significances are listed in Table~\ref{table:table2}. The black lines with error bars in each panel represents all central-satellite pairs in our entire sample. Clearly, when we consider all galaxy pairs, the dependence of large-scale environments on the SCA is fairly weak. Note however that there is still a tendency for the alignment angle to increase from voids to walls, then to filaments, while slightly decreasing in knots.  Note that, this is the mixed effect of the number fraction of blue and red centrals in different LSE. That will be discussed later.

For galaxy groups defined by red centrals, (red line in the upper-left panel, the red and green lines at the bottom-left panel), there is no effect of the LSE on the SCA. When only considering the colour of satellites which are shown at the upper-right panel in Fig.~\ref{fig:fig1}, the environment dependence is still weak, showing trends similar to the ``All-sample''. However, for those satellite-central pairs with blue central galaxies (blue line in the upper-left panel, the blue and green lines in the bottom-right panel in Fig.~\ref{fig:fig1}), we find the alignment angle is strongly dependent on the LSE. The alignment angle increases from void to wall to filament to knot environment for groups with blue centrals, independent of satellite colour. This suggests that the denser the environment of a blue central galaxy, the weaker the SCA of the group is. What more interesting is that for those satellites in the knot environment, the alignment angle is larger than $45^{\circ}$ with $3.7\sigma$ significance. This means those satellites tend to be perpendicular to the major axis of central galaxy, (reminiscent of the $Holmberg$ effect).  In addition, at a fixed LSE, the mean alignment angle for 'red centrals' sample is always smaller than 'blue centrals' sample. Similarly, red satellites are more preferentially aligned with the major axis of central galaxy than blue satellites in general (the red lines in the upper-right and bottom-left panel), which agrees with previous works and Table~\ref{table:table1}.

\begin{figure*}
\centerline{\includegraphics[width=0.85\textwidth]{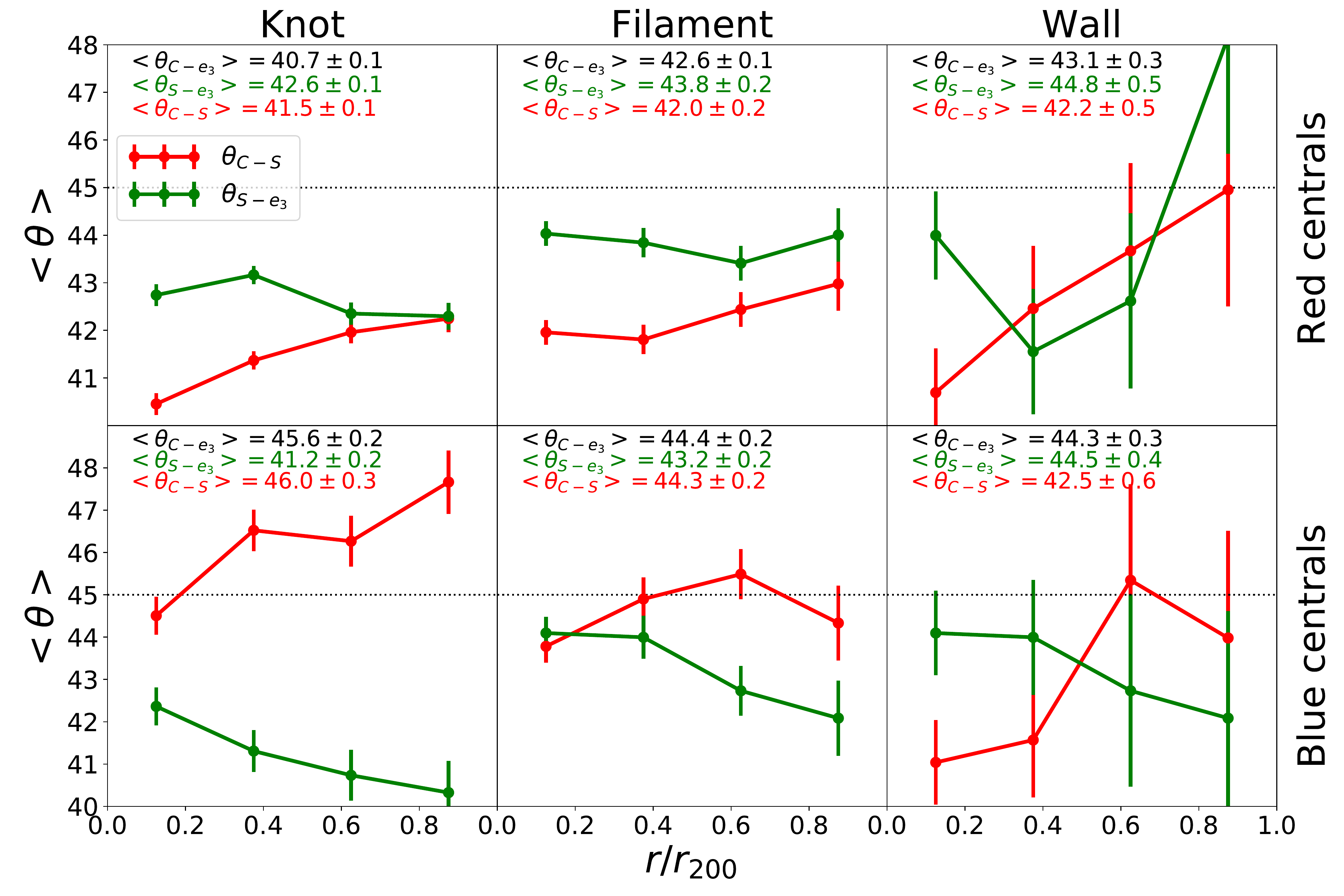}}
\caption{The mean angle $\theta$, as indicated in the legend, as function of the projected radius, $r$, in units of the virial radius of the host group, $r_{200}$,  between the satellite galaxies and the central galaxies considering the colour of central galaxies (upper panels for red centrals and bottom panels for blue centrals) and the environment of groups (as indicated at the top of each column).  Here $\theta_{\rm C-S}$ indicated the angle between central and satellites (red solid lines), $\theta_{\rm S-e_3}$ indicated the angle between the vector of satellites-central and the $\eee$ of LSS (green dashed lines).  At the top of each panel, we list the mean value of $\theta_{C-S}$, $\theta_{S-e_{3}}$ and the angle $\theta_{\rm C-e_3}$ which represents the angle between the major axis of central galaxy and the $\eee$ of LSS.}
\label{fig:fig3}
\end{figure*}

Next, in order to check the colour dependence of the SCA in different LSE, we investigate the alignment angle as a function of the ${}^{0.1}{(g-r)}$ colour of central galaxies in knots, filaments and walls (we ignore voids due to the small sample size). In the left panel of the Fig.~\ref{fig:fig2}, we show the distribution of the colour of central galaxies in our sample. It can be found that, at the low values of the ${}^{0.1}{(g-r)}$, most central galaxies are in walls and the fraction of central galaxies in knots is the lowest. Whereas for high values of ${}^{0.1}{(g-r)}$, the peak value of the colour distribution increases from walls to knots. The distribution of central galaxies within different LSE means that most of the red central galaxies are located in the relatively high density LSE (filaments and knots), and most of the blue central galaxies appear in the relatively low dense LSE (walls). This is the main reason for the trend seen in the "All-sample" in Fig.~\ref{fig:fig1}: in knot and filament environment the SCA of "All-sample" is close to the "red centrals" sample, while in wall and void environment the SCA of "All-sample" is close to the "blue centrals" sample. In the right panel of  Fig.~\ref{fig:fig2}, it can be seen that the $\theta_{C-S}$ generally decreases with increasing central galaxy colour index. The redder the central galaxy is,  the stronger of the SCA is, independent of LSE. In walls and filaments, the SCA angle is always smaller than $45^{\circ}$, which means that the satellites always tend to align with the major axis of their central galaxy. However, for satellites hosted by the low ${}^{0.1}{(g-r)}$ central galaxy in knots, the SCA angle is larger than $45^{\circ}$. This indicates that those satellites tend to perpendicular to the major axis of central galaxy, which agrees with the case of blue central galaxy dominated groups in Fig.~\ref{fig:fig2}. For satellites hosted by the same ${}^{0.1}{(g-r)}$ colour of central galaxy, their angular distribution around the major axis of central galaxy is affected by the LSE. In other words, the LSE can not be ignored when examining the satellites distribution in a group.

\begin{figure*}
\centerline{\includegraphics[width=0.85\textwidth]{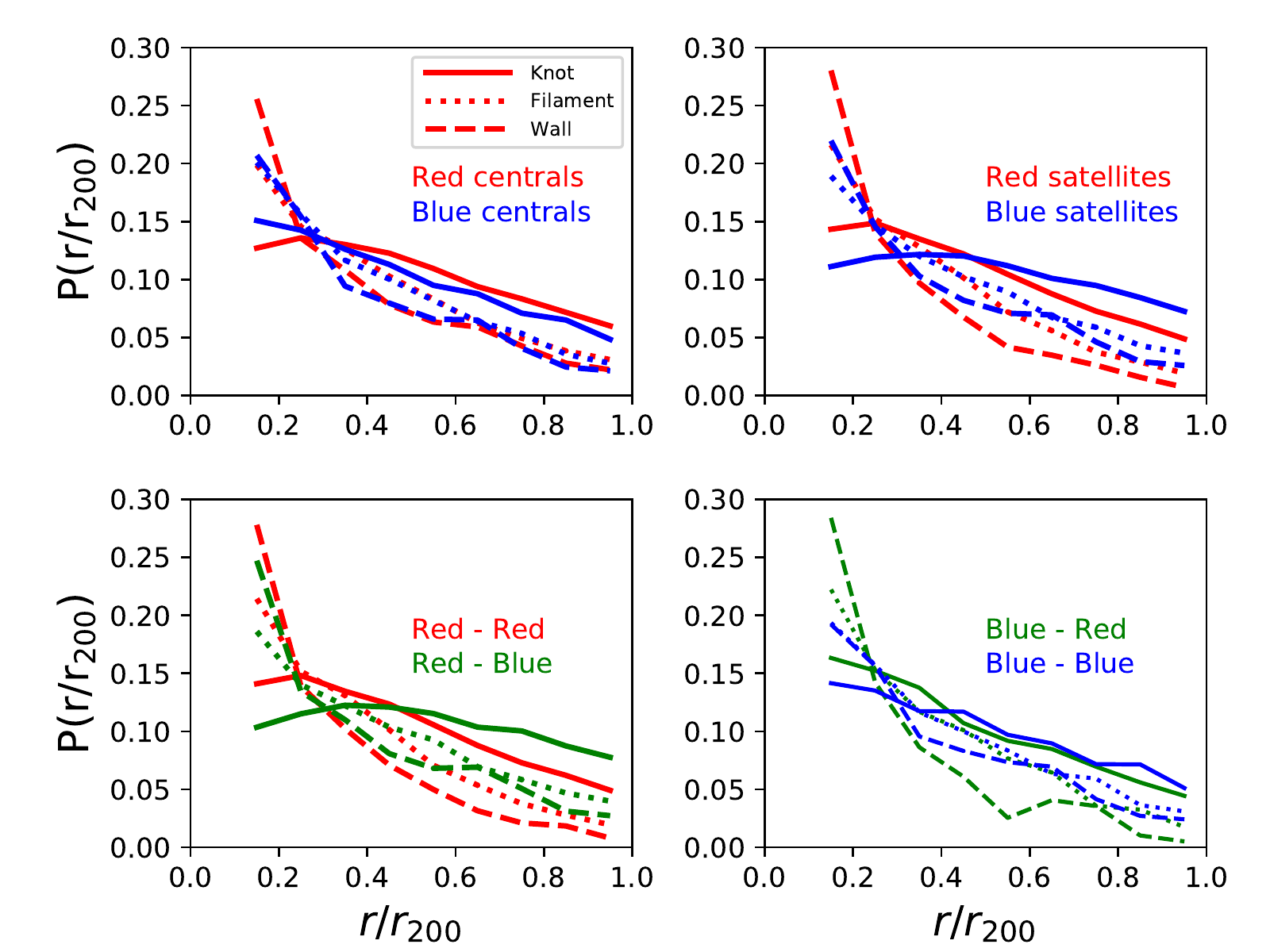}}
\caption{The satellites radial distribution in knot, filament and wall of each sub-sample. The solid, dashed and dotted lines represent galaxies in knots, filaments and walls, respectively.}
\label{fig:fig4}
\end{figure*}

\subsection{Radial distribution of the SCA}

In above subsections, we study the SCA by examining the mean alignment angle of all satellites in a group. However, as mentioned in the introduction, dark matter is believed to be anisotropically distributed in the halo \citep{2015MNRAS.450.2727T, 2016ApJ...830..121L}. This compels us to ask the question {\it is the SCA angle different when computed at different distances? And if so could this difference  be related to the LSE and galaxy colour?} In order to address this, we investigate the dependence of three different alignment angles: $\theta_{\rm C-S}$, $\theta_{\rm C-e_3}$ and $\theta_{\rm S-e_3}$ (their definitions can be found in Section 2.3) in the different LSE: knots, filaments and walls. We define the satellite's radial position as $r/r_{200}$, namely the projected distance to the group center scaled by virial radius.

In Fig.~\ref{fig:fig3}, we show the mean angle formed between the central galaxy's major axis and $\eee$ ($\langle\theta_{\rm C-e_3}\rangle$) and that formed between the satellites position and $\eee$ ($\langle\theta_{\rm S-e_3}\rangle$) as a function of the satellites radial position. We further examine this alignment as function of both central colour (rows) and LSE (columns). In the top (bottom) three panels we show the mean angle for groups with red (blue) centrals. Each column shows groups in different LSEs, namely knots, filaments and walls as indicated at the top of each column. At the top of each panel, we print the mean value of the angles averaged over the entire radial extent. It can be seen that, from walls to filaments then to knots, the $\langle\theta_{\rm C-e_3}\rangle$ decreases for red central galaxies and increases for blue central galaxies, meanwhile the $\langle\theta_{\rm S-e_3}\rangle$ decreases irrespective of central galaxy colour. Even at a given LSE, the $\langle\theta_{\rm C-e_3}\rangle$  for red centrals is systematically smaller than 'Blue centrals', which means the red central galaxies are more likely to align with $\eee$ direction than blue central galaxies. This result agrees with that found by \cite{2009ApJ...706..747Z}, \citep[see also][]{2015MNRAS.450.2727T}, however in that work, they only examined the $\langle\theta_{\rm C-e_3}\rangle$ in filaments and walls. It also can be seen that even at inner part ($\rm r/r_{200}\sim0.1$) of blue central, $\theta_{\rm C-S}$ still has LSE dependence which the alignment angle increases from wall to knot.

The $\langle\theta_{\rm S-e_3}\rangle$ (green line in each panel) shows different trends with increasing radius in different LSE. For red centrals in knots and filaments, $\langle\theta_{\rm S-e_3}\rangle$ is almost independent of satellite distance. For blue centrals however, $\langle\theta_{\rm S-e_3}\rangle$ clearly   decreases with radius. It also can be seen that in a given LSE: both knots and filaments, the $\langle\theta_{\rm S-e_3}\rangle$ for blue central galaxy groups is overall smaller than it for red central galaxy groups at the same radius. While in walls, $\langle\theta_{\rm S-e_3}\rangle$ the radial dependence of the SCA for both red and blue central galaxies samples are quite different. For blue central galaxy groups, $\langle\theta_{\rm S-e_3}\rangle$ is smaller than $45^{\circ}$  at whole range radius. For red central galaxy groups, $\langle\theta_{\rm S-e_3}\rangle$ is smaller than $45^{\circ}$ at the inner region, whereas $\langle\theta_{\rm S-e_3}\rangle$ is greater than $45^{\circ}$ at the outer region. The angle between satellite position and $\eee$ reveals the effect of a group being accreted along the direction defined by the LSS and satellite orbit evolution within a group. Generally, the smaller $\langle\theta_{\rm S-e_3}\rangle$ in both knots and filaments for both red and blue centrals, means the satellites these dense LSE are better aligned with the accretion direction. This is especially for blue centrals. We can thus infer that the distribution of satellites in the densest LSE almost follows the $\eee$ direction. \textit{When compared with blue centrals, the relatively flat $\langle\theta_{\rm S-e_3}\rangle$ radial distribution of red centrals implies that experiencing more merger events may dominate the satellites distribution}.

We also found that the mean angle of SCA, $\theta_{\rm C-S}$ (red lines in each panel), increases with increasing radius, and the $\theta_{\rm C-S}$ is always smaller than $45^{\circ}$ for red centrals in these three LSE. This means satellites in groups with red centrals tend to align with the major axis of central galaxy, and the strength of alignment decreases (the $\theta_{\rm C-S}$ increased) with increasing radius. This is consistent with Y06. But for blue centrals in knots, filaments and walls, it becomes larger than $45^{\circ}$ at some radius. For blue centrals in knots, their satellites tend to be perpendicular to the major axis at $r/r_{200} > 0.2$, and the strength of this perpendicular alignment increases with increasing radius. Similarly, in filaments and walls, the perpendicular signal appears at outer part of host groups. We checked that if we do not consider the environment of blue centrals, the $\langle\theta_{\rm C-S}\rangle$ is absent with projected radius and consistent with Y06 as well.
Another interesting result from Fig.~\ref{fig:fig3} is that in the inner region, the SCA of blue centrals decreases with the density of LSE increases (from walls to knots), while the
SCA of red central roughly increases (except in walls). This also indicates that the inner satellites may be more affected by the central galaxies via mergers or tidal forces.

We futher examine the distributions of the radial number fraction of satellites with different colours in different LSE in Fig.~\ref{fig:fig4} to
study the role central galaxies play in SCA. It can be seen, in general, the intersections of these lines for each sub-sample are located between $r/r_{200}=0.2$ to $r/r_{200}=0.4$, and with a mean value of $r/r_{200}=0.25$. Satellites (especially red satellites) prefer to reside in the inner part of groups and the number fraction decreases with the increasing radius. Such evolutionary behaviour is consistent with ``strangulation'' wherein cold gas that could be used for star formation is depleted as satellites are accreted \citep[e.g.,][]{2004MNRAS.353..713K, 2005ApJ...631...21K, 2016MNRAS.458..366L}. The shapes of each curve (the maximum value, the decrease gradient) not only depends on the colour of central and satellite galaxies but also depends on the LSE.  In the inner part of haloes, the fraction of satellites is lower in the more dense LSEs. In the outer parts, it's the opposite namely in knots there is a higher fraction of satellites. It need to be pointed out is that, in a given environment and for a given central colour, the fraction of blue satellites is always higher than red satellites in outer regions. For wall environments, about $25\%\sim30\%$ satellites reside at the low values of $r/r_{200}$, and the fraction decreases very quickly. Additionally, in the inner parts, the fraction of satellites of red centrals (red dashed line in top left panel of Fig.~\ref{fig:fig4}) is $\sim5\%$ higher than those in blue centrals (blue dashed line in top left panel of Fig.~\ref{fig:fig4}). The radial distributions of satellites in filaments are almost absent with the different colours of both central and satellite galaxies. However, in a knot environments, blue centrals have $\sim3\%$ more satellites than red centrals at low values of projected radius.
These radial satellites distributions with different colours in different LSE indicate that, red central galaxies have merged more satellites than blue
central galaxies in the knot LSE, which means satellites with a red central galaxy may be more affected by the interaction with central galaxies  and represent a better SCA especially in the inner region.


\section{Discussion}

The main finding of this work is a strong large scale environment dependence of the satellite central alignment (SCA) for blue central galaxies which suggests that the LSEs are needed to be considered in investigations of the satellite distribution, such as the satellites system in local group. The mis-alignment case of satellites hosted by blue centrals is similar to the $Holmberg$ effect and also related to the satellites system distribution of our Milky Way, M31 and Centaurus A.  Combining the radial distribution of satellites in different environments and the radial dependence of the  $\theta_{\rm S-e_3}$, we can speculate the LSE dependence of satellites angular distribution around the major axis of the central galaxy.  In this section, we will discuss the possible origin of this satellite-central alignment within different LSEs.

According to the current models of galaxy formation \citep[e.g.,][]{2008ApJ...676L.101K, 2011MNRAS.413..101G}, it is expected that early accreted sub-haloes (satellites) often reside in the inner region of the host halo (group) and mostly being red, and the later accreted sub-haloes (satellites) should orbit at the outer region and mostly being blue. This can also be seen in our results (the right top panel of Fig.~\ref{fig:fig4}). In general, those red satellites have suffered the tidal field of central galaxies for longer time, and have lost the LSS information. So these red satellites follow more closely the shape of the host halo in inner region, giving rise to stronger satellite-central alignment, which can be seen from Fig.~\ref{fig:fig3}.

For groups with blue central galaxies, according to the results in Fig.~\ref{fig:fig3} and Fig.~\ref{fig:fig4}, the LSE dependence can be explained from the following two facts. On one hand, from the wall to knot environment, the angle ($\rm \theta_{C-e_{3}}$, the alignment between the major axis of central galaxy with the large-scale e3) is almost randomly distributed (Fig.~\ref{fig:fig3}).  Although the infall satellites carry some alignment with e3, on average the infall of satellites is not aligned with the central galaxy. On the other hand, from wall to knot environment, there are more satellites are being accreted, with more satellite galaxies are distributed in outer halo region (Fig.~\ref{fig:fig4}). Compared to the isolated wall environment, the strong infall of satellites (but are randomly distributed relative to central galaxy) in dense knot environment dilutes the alignment signal established by tidal force of central galaxy with its satellites. So basically, the strong environment dependence of SCA is a competition between halo internal evolution and the infall of satellites on large scales.

For the alignment in groups with red central galaxy, the SCA with environment is weak. This is also a consequence of the competition between halo internal evolution and infall of satellite on large scales. From galaxy formation model, it is well known that red galaxy is often the remnant of galaxy merger \citep[e.g.,][]{2005ApJ...631...21K}. After the accreted satellite galaxies have merged with the central galaxy, the central galaxy will be more likely an elliptical and being red. As the accreted satellites carry more alignment with $\eee$, so the shape of central galaxy will be aligned with $\eee$ \citep[Fig.~\ref{fig:fig3}, also in][]{2013ApJ...779..160Z} with stronger alignment in knots than in wall. But in wall environment the satellites are more likely distributed in inner halo region, leading to a stronger alignment between central and satellites. The combined effects on stronger C-$\eee$ alignment in knots and stronger SCA in halo inner region lead to a weaker dependence on environment for groups with red central galaxy.

To summarize, the dependence of the SCA on the colour of the central galaxies and the LSE arises from the competition between the effect of satellite accretion on large-scale and the internal evolution effect in the halo (the tidal force and merger of satellites with central). We note that such a scenario is the best model which can explain the overall results seen in our work and it could be checked in future by using hydro-dynamical simulation which can trace the evolution of the galaxy.

\section{conclusion }

Using the catalogue constructed from SDSS DR7, we investigate the LSE dependence of the alignment between the position vector of satellites and the major axis of central galaxies, the alignment among the satellites, centrals and $\eee$ of LSS, the radial distribution of satellites within LSS. The main results are summarized as follow:

\begin{itemize}
\item In agreement with previous studies \cite[e.g.,][]{2006MNRAS.369.1293Y}, we obtained the same  signal (but higher significance given our larger sample size) that, considering all satellite-central
pairs, satellites are preferentially distribution, for a mean value $42.2^{\circ}$, around the orientation of the major axis of the central galaxy. The
strongest alignment appears between red satellites and red central galaxies in most massive groups.

\item We find that the satellite-central alignment strongly depend on LSE for blue central galaxies dominated groups, and is absent for groups with red central galaxies. In groups with the blue central galaxies, the alignment angle decreases with the LSE order: knots, filaments, walls, voids. For a knot environment, satellites tend to be perpendicular to the major axis of its blue central slightly (the alignment angle is approximately $46^{\circ}$). Whereas for red centrals, the alignment angle drops to $\sim41^{\circ}$ to show an parallel alignment signal at all type of LSE.

\item After investigating the angles among satellites, centrals and $\eee$, as well as the radial distribution within different LSE, we find that more satellites  preferentially reside in the inner part of halo in wall, and a much weaker radial distribution in knots.  Combined with previous works about sub-haloes accretion and mass accretion of dark matter haloes \citep{2011MNRAS.411.1525L, 2014MNRAS.443.1274L, 2015ApJ...807...37S, 2015ApJ...813....6K,2018MNRAS.473.1562W} , We found the the satellite-central alignment is the competition between the satellite accretion within LSE and galaxy evolution inside haloes. For groups contain red central galaxies, the satellite-central alignment is mainly contributed by the evolution effect, while for blue central dominated groups, the effect of large scale structure plays a more important role, especially in knots environments.

\end{itemize}


\acknowledgments
{\bf Acknowledgement:}
We thank the anonymous referee for comments that substantially improve the manuscript. The author (PW) thanks Feng Shi (SHAO) and Ian Roberts (McMaster University) for help to calculate the projected distance between satellites and centrals. We thanks Xiaohu Yang for making the group catalogue of SDSS DR7 public available, and Huiyuan Wang for calculation the large scale structure used in this work. We also thanks Shi Shao (Durham) and Marius Cautun (Durham) for valuable suggestions and discussions. The work  is supported  by the NSFC (No.  11333008), the  973 program  (No.  2015CB857003, No.  2013CB834900), the NSFC (No.11703091), the NSF of Jiangsu Province (No. BK20140050). QG acknowledges the support from the NSFC (no. 11743003 ). ET was supported by ETAg grants (IUT40-2, IUT26-2) and by EU through the ERDF CoE grant TK133.

\bibliographystyle{apj}

\end{document}